\documentstyle[prl,aps,twocolumn,epsfig]{revtex}
\begin{document}

\draft
\title{Dynamic Scaling at the Zero-field 2D Superconducting Transition}

\twocolumn[ 
\hsize\textwidth\columnwidth\hsize\csname@twocolumnfalse\endcsname 

\author{S.~M.~Ammirata,$^1$ Mark Friesen,$^2$ Stephen 
W.~Pierson,$^3$ 
LeRoy A.~Gorham,$^3$ Jeffrey C. Hunnicutt,$^3$ M.~L.~Trawick,$^1$ 
and C.~D.~Keener$^1$}
\address{$^1$Department of Physics, Ohio State University, Columbus, 
Ohio 43210}  
\address{$^2$Physics Department, Purdue University, West Lafayette, IN 
47907-1396}  
\address{$^3$Department of Physics, Worcester Polytechnic Institute 
(WPI), 
Worcester, MA 01609-2280}    

\date{\today}
\maketitle

\begin{abstract} 
Zero-field current-voltage ($I$-$V$) characteristics of a thin 
(``two-dimensional'') 
Bi$_2$Sr$_2$CaCu$_2$O$_{8+\delta}$ crystal are reported and 
analyzed in two ways. The 
``conventional'' approach yields ambiguous results while a dynamical 
scaling analysis offers new 
insights into the Kosterlitz-Thouless-Berezinskii transition. 
The scaling theory predicts that the universal jump
of the $I$-$V$ exponent $\alpha$ 
should be between $z+1$ and 1.  
A value of $z\simeq 5.6$ is obtained for the dynamical critical exponent, 
and is corroborated by data 
from other 2D superconductors.
A simple dynamical model is presented to account for the results.

\end{abstract} 
]
\narrowtext

Perhaps no transition is better known than the 
Kos\-ter\-litz-Thouless-Berezinskii (KTB) 
transition,\cite{ktb} which occurs in systems in the universality class of the 
two-dimensional (2D) 
$XY$ model and which is marked by the unbinding of topological 
excitations known as vortex 
pairs. Since the original theory was derived in the early 1970's, the 
KTB transition has been 
mentioned in systems as diverse as superfluids,\cite{br78} 
superconductors,\cite{hn79} 2D lattices,\cite{huberman79} 
ferromagnets,\cite{mertens87} and liquid crystals.\cite{geer89}

Evidence for Kosterlitz-Thouless-Berezinskii (KTB) critical behavior in 
two-dimensional 
superconductors in zero-field was first reported\cite{epstein81} in 
1981 and has since been widely studied in 
thin films,\cite{hebard83,goldman83} 2D Josephson Junction 
arrays,\cite{resnick81} and more 
recently in high-temperature superconductor (HTSC) ultrathin 
films.\cite{norton93} 
Evidence for KTB behavior (but not necessarily a transition) also abounds 
for multilayers 
and HTSC single crystals.\cite{ktexp} 

Two main approaches can be used to investigate electronic transport
phenomena in superconductors near the KTB transition.
In the first, more ``conventional'' approach,
there are two principal signatures of KTB behavior.\cite{hn79,minnhagen87}  
First, the ohmic resistance
should have a unique temperature dependence above the superconducting
transition temperature, $T_{KTB}$:
$R(T) \propto \exp[-2\sqrt{b/(T/T_{KTB}-1)}]$, where
$b$ is a non-universal constant.  Second, 
current-voltage ($I$-$V$) isotherms should be described by a
power law, $V\propto I^\alpha$, at low currents.\cite{hn79}  Below $T_{KTB}$,
$\alpha$ decreases linearly with increasing temperature,
makes a ``universal jump''
from three to one at $T=T_{KTB}$,\cite{nelson77} then remains ohmic 
($\alpha =1$) for all $T>T_{KTB}$.

There are several difficulties with this approach.
For example, the jump in $\alpha$ is strictly defined only 
in the $I\rightarrow 0$ limit, making it difficult to detect experimentally. 
Indeed, many of the original 
papers\cite{epstein81,hebard83,goldman83,resnick81} 
do not observe or report universal jumps.  
Furthermore, it cannot address the 
eventual upturn of ohmic isotherms, observed near $T_{KTB}$.\cite{lobb96}  
(See Fig.~1 below, and Refs.~\cite{hebard83,goldman83,lobb96,rutgers91}.) 
Similar behavior is also observed in finite field near the superconducting 
transition 
in bulk superconductors, where it is attributed to the competition between
thermal and current-induced effects.\cite{blatter94,ffh}  In the second
approach, in which one applies the techniques
developed for such bulk systems, it is possible to resolve both of these
issues.

The second approach treats {\it both} low and high current behaviors, via
dynamic scaling.  The main ideas were 
first discussed for 3D~$XY$ and glassy critical phenomena
in bulk superconductors.\cite{ffh}  In this analysis, 
dynamic phenomena near the critical point
involves a correlation time scale $\tau \propto \xi^z$, which like the
correlation length $\xi$, diverges at $T_{KTB}$.  ($z$ is 
the dynamic critical exponent.)
The scaling approach has two distinct advantages 
over the conventional theory.  
First, it extends the transport theory to finite currents, and is
able to describe the observed upturn of ohmic isotherms.
Second, it admits dynamics which are non-diffusive.
This issue has recently received much attention 
in bulk HTSC's near the 3D~$XY$ critical
point, where unexpected (non-diffusive) dynamics have been
encountered.\cite{booth} 

For 2D superconductors, the $I$-$V$ curves should scale as
$V=I\xi^{-z} \chi_\pm(I\xi/T)$,\cite{ffh,wolf79} 
where $\chi_{+ (-)} (x)$ is the scaling function for temperatures 
above (below) $T_{KTB}$.  
The asymptotic behaviors of $\chi_\pm (x)$ can be deduced from the 
conventional KTB theory:  
$\lim_{x\rightarrow 0} \chi_+ (x)= \mbox{const.}$ (ohmic limit);
$\lim_{x\rightarrow 0} \chi_- (x)\propto \exp{[-a\ln (1/x)]}$
(thermally activated vortex pair unbinding\cite{note1});
$\lim_{x\rightarrow \infty}\chi_\pm (x)\propto x^z$ (critical isotherm).
The univeral jump is expressed as the difference in (log-log) slopes 
between the two asymptotic limits of $\chi_+(x)$. 
The distinctions between the scaling 
and the conventional approaches are apparent:
(i) $R(T) \propto \exp[-z\sqrt{b/(T/T_{KTB}-1)}]$ 
($2$ replaced by $z$); and (ii) the ``universal'' jump of
$\alpha$ is now from $z+1$ to 1, not from 3 to 1. Of 
course the two approaches are equivalent for diffusive dynamics ($z=2$). 
Despite the recent widespread application of scaling techniques to 
$I$-$V$ characteristics near a critical transition,\cite{ffh}
we are familiar with only two cases where these ideas have been
applied to 2D superconductors.\cite{wolf79,miu95} In the former, the dynamic 
universality class was not explicitly studied while in the latter, the issue
was treated only peripherally.

In this Letter, we apply both approaches to $I$-$V$ curves from a 
thin crystal of Bi$_2$Sr$_2$CaCu$_2$O$_{8+\delta}$ (BSCCO). 
It is found that the conventional approach 
gives rather ambiguous results.  However, by applying dynamical
scaling, with $z$ as a fitting parameter, the data can be made to 
collapse, obtaining $z\simeq 5.6\pm0.3$ ({\it not} $z=2$).  To check 
the universality of our results, 
we also apply the scaling analysis to other $I$-$V$ data sets
obtained from the literature, including both HTSC and 
conventional 2D superconductors, with an average result of $z\simeq 5.7$.

BSCCO crystals were prepared using a standard self flux 
method.  A single crystal was selected and cleaved using a 
Scotch tape
technique.\cite{trawick96}  Thin $1500~\AA$ crystals were obtained 
with atomically smooth surface areas of at least 
(50 micron)$^2$.\cite{keener97a} Thicknesses were measured using an 
atomic force microscope.
Four silver pads were photolithographically patterned on top of the crystal.
The crystals were 
postannealed at 600 C in 1.5 scfh flowing O$_2$ for 45 minutes to achieve
optimum oxygenation. Gold wires were 
ultrasonically bonded to the silver pads where they extended beyond
the crystal. 

$R(T)$ and $I$-$V$ characteristics were
measured using a standard four probe inline geometry and a low frequency
lock-in technique (16.9 Hz). A temperature
controller was used to insure a temperature stability
of better than 10 mK. 
Although current was injected only 
through the top crystal surface, the thickness of the sample assured uniform 
current distribution. 
Data was taken on three different samples. 
Data presented here is from a crystal of dimensions 0.2mm x 0.5mm x 
($1500 \pm 500~\AA$).  For thicker samples, KTB behavior could not 
be observed.

\begin{figure}
\centerline{
\epsfig{file=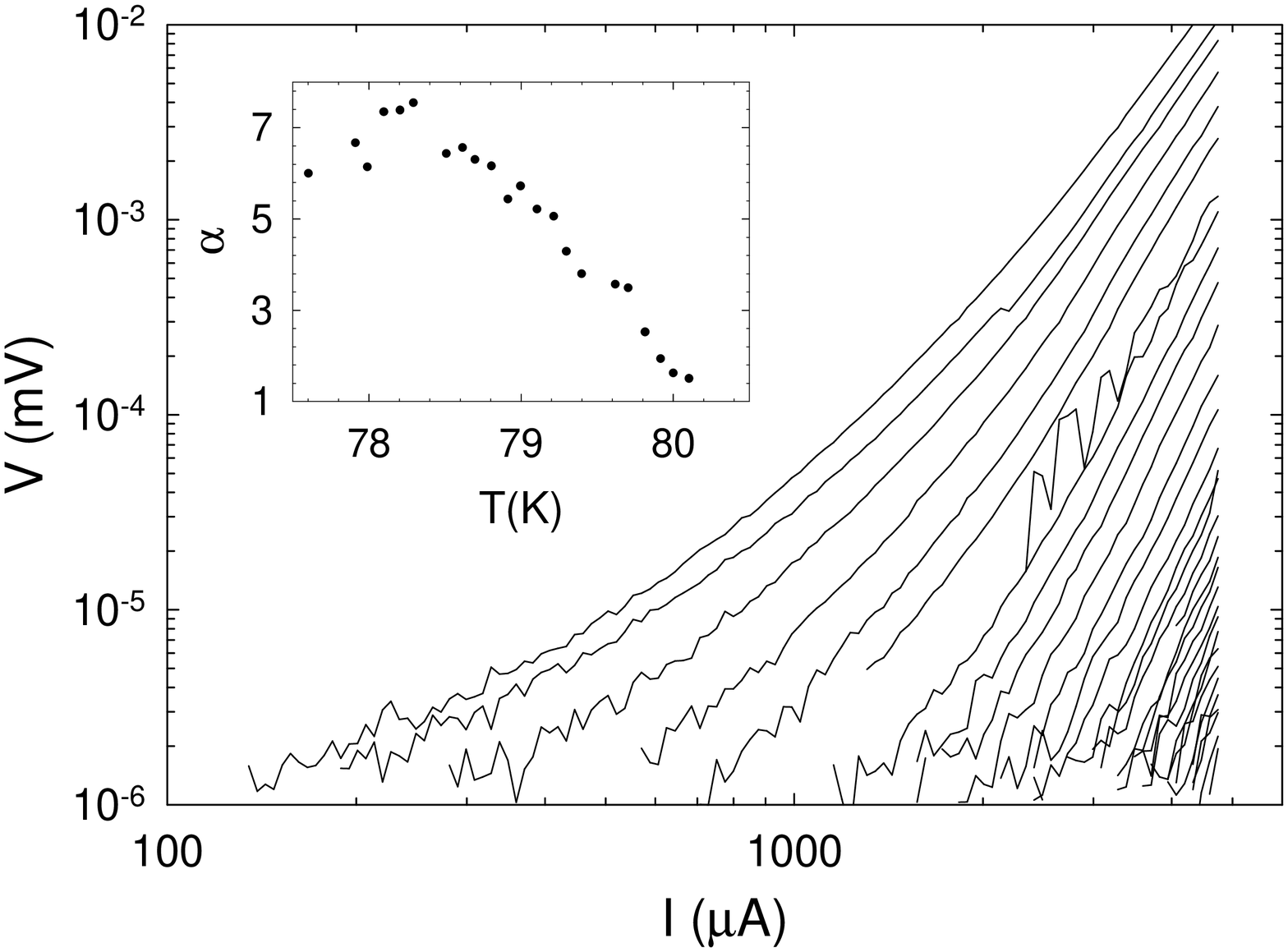,width=3.4in}}
\caption{$I$-$V$ isotherms from a thin
Bi$_2$Sr$_2$CaCu$_2$O$_{8+\delta}$
crystal for temperatures 76.9 K to 80.1 K. [INSET:
$\alpha(T)$ determined by a fit to $V\propto I^\alpha$ in the voltage
range 2 to 10 nV.]}
\label{sergiodata}
\end{figure}

Fig.~\ref{sergiodata} shows zero field ($H\lesssim$~100~mGauss) 
$I$-$V$ isotherms, corresponding to 100 mK steps, 
from $T = 80.1$ K (left side) to $76.9$ K (right side). 
The curves display features typical of 2D samples.  The 
high temperature ($>T_{KTB}$) isotherms exhibit positive curvature
as they cross over from ohmic to non-ohmic behavior,
while the low temperature ($<T_{KTB}$) isotherms 
exhibit negative curvature.  The critical 
isotherm separates the two regions at $T=T_{KTB}$, and describes a strict power
law:  $V\propto I^{1+z}$.  In Fig.~\ref{sergiodata}, and for other published
2D isotherms ({\it e.g.}, Refs.~\cite{goldman83,lobb96,rutgers91}),
identification of the critical isotherm 
shows that $z$ is clearly larger than 2.

We first perform a conventional analysis.  The ohmic resistance
data $R(T)$ is fit to the KTB prediction, with apparent success, obtaining
$T_{KTB}=78.87$~K and $b=1.3$.  The $I$-$V$ exponent $\alpha (T)$
is then determined by fitting $V\propto I^\alpha$ to the lower portion 
(2 to 10 nV) of each isotherm.  
In Fig.~\ref{sergiodata}, high temperature (low current)
ohmic behavior, expected to persist for all $T>T_{KTB}$, is obscured by
noise already for $T<79.7$~K.  Such experimental
limitations lead to significant errors in
$\alpha (T)$ near and below $T_{KTB}$, thus smearing out the universal jump
until it becomes unrecognizable, as in Fig.~\ref{sergiodata} (inset).  These
same difficulties affect $R(T)$ measurements, reducing the accuracy of
the obtained fitting parameters.  

In the dynamic scaling analysis, we are not
restricted to low currents, so
that voltage sensitivity no longer becomes an issue.
It is convenient to rewrite the scaling ansatz as\cite{SCdetail} 
\begin{equation}
I^{z+1}/[VT^z]=\varepsilon_\pm(I^z\xi^z/T^z)
\label{FFHsc}
\end{equation}
where $\varepsilon_\pm (x)\equiv x/\chi_\pm (x^{1/z})$.  
Note that above $T_{KTB}$, the scaling variable can also be written as
$x=I^z/R(T)T^z$.  To proceed 
with the analysis, $\xi (T)$ must
be specified.  We will assume that the KTB form,
$\xi (T)\propto \exp[\sqrt{b/(T/T_{KTB}-1)}]$, provides the most
efficient parameterization of the correlation length. We further assume 
that $\xi (T)$ is symmetric about $T=T_{KTB}$
(modulo some prefactor).\cite{symmdetail}
The fitting parameters for Eq.~(\ref{FFHsc}) then become $z$ (universal),
and $T_{KTB}$ and $b$ (nonuniversal).  The following three requirements
must be fulfilled self-consistently in our scaling procedure:
(i) $V\propto I^{z+1}$, along the critical isotherm $T=T_{KTB}$;
(ii) $R(T)\propto \xi^{-z}$, in the high temperature range ($\gtrsim 79.7$~K),
where ohmic $R(T)$ can be obtained; and
(iii) scaling collapse of the $I$-$V$ isotherms, according to 
Eq.~(\ref{FFHsc}). The first two points place tight constraints on
the fitting parameters.  Within these constraints, the parameters are tuned
to satisfy the final point.

\begin{figure}
\centerline{
\epsfig{file=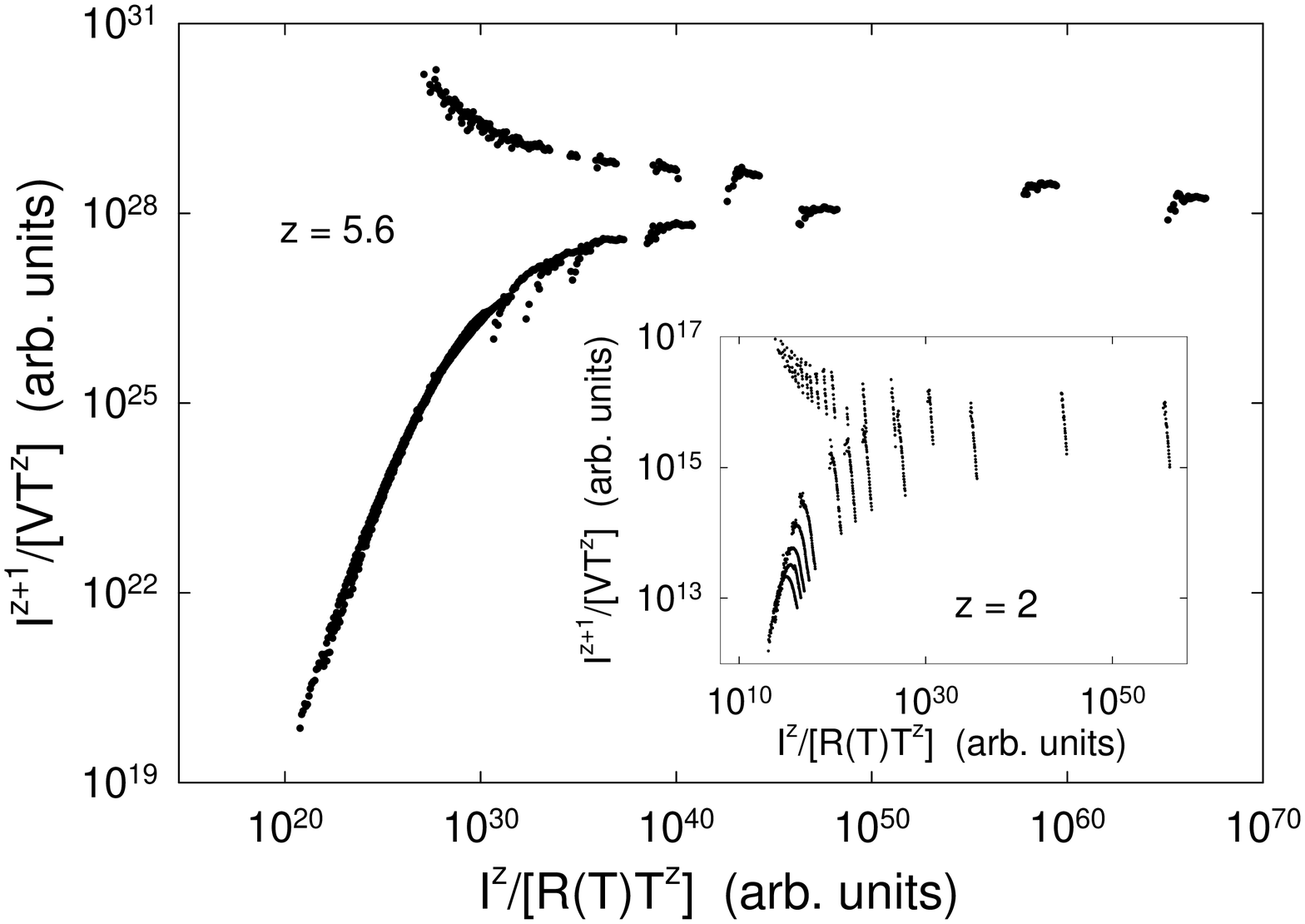,width=3.4in}}
\caption{The $I$-$V$ curves of Fig.~\protect\ref{sergiodata}
scaled with Eq.~(\protect\ref{FFHsc}). [INSET: The same scaling for $z=2$.]}
\label{sergiosc}
\end{figure}

\begin{figure}
\centerline{
\epsfig{file=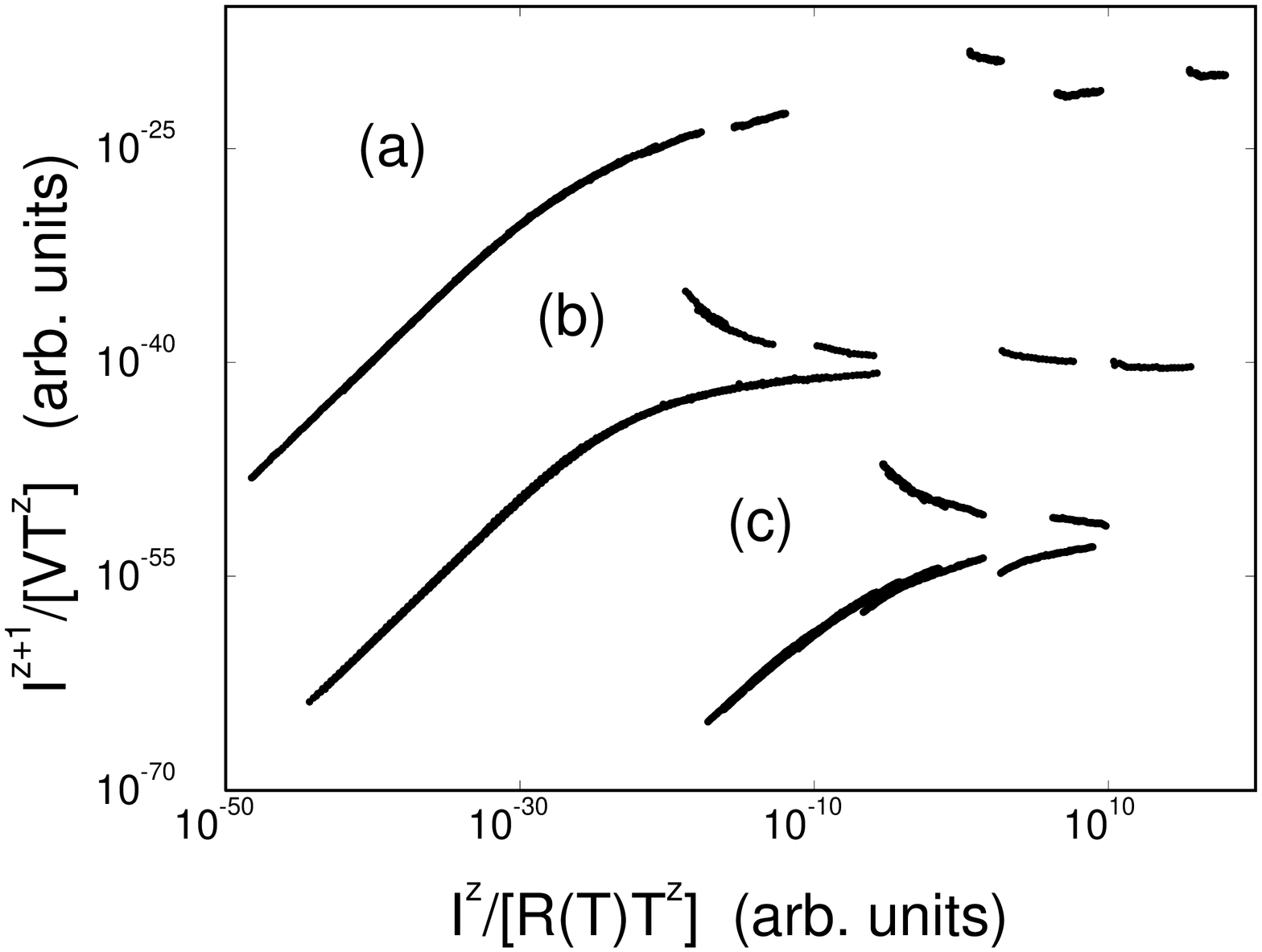,angle=-90,width=3.25in}}
\vskip2mm
\caption{The $I$-$V$ curves of (a)
Ref.~\protect\onlinecite{rutgers91} (b) Ref.~\protect\onlinecite{lobb96}
and (c) Ref.~\protect\onlinecite{goldman83} scaled with
Eq.~(\protect\ref{FFHsc}).
Data sets (b) and (c) have been shifted arbitrarily.}
\label{repacisc}
\end{figure}

The outcome of such a fitting procedure for the data of Fig.~\ref{sergiodata}
gives the following results:  $T_{KTB}=78.87$K, 
$b =0.57$, and $z=5.6\pm0.3$.  The scaling collapse,
shown in Fig.~(\ref{sergiosc}), is convincing in both the upper 
($T<T_{KTB}$) and lower ($T>T_{KTB}$) branches.  For many isotherms
in Fig.~\ref{sergiosc}, the 
noisy (low current) portions of the curves do not scale; this may be due
to electromagnetic screening (finite size effects).\cite{lobb96} 
However the same 
difficulty is not incurred for the other data
analyzed below.  (See Fig.~\ref{repacisc}.)  The caution expressed earlier
for the low temperature branch\cite{note1} does not seem to affect
the scaling collapse.  The inset of Fig.~\ref{sergiosc}
shows the results of a similar two-parameter fit to the data, obtained
by setting $z=2$. The results are extremely poor, indicating that $z>2$.

To check our results, we have applied 
the same scaling procedure to other data 
from the literature. The first data set\cite{rutgers91} is from a 
YBCO/PrBa$_2$Cu$_3$O$_{7-\delta}$ multilayer in which the YBCO 
layers have a thickness of $24~\AA$ and are electrically  isolated from one 
another by PrBa$_2$Cu$_3$O$_{7-\delta}$ barrier layers.  The scaling
procedure now leads to the results  
$T_{KTB}=32.0$~K, $b =39.3$, and $z=5.6\pm0.3$, with the $I$-$V$ collapse
shown in Fig.~\ref{repacisc}a.
Analysis of $I$-$V$ data\cite{lobb96} from a $12~\AA$ YBCO monolayer 
gives $T_{KTB}=18.5$~K, $b=22.29$, and $z=5.8\pm0.3$ and also  
collapses the data. (See Fig.~\ref{repacisc}b.) The final data set 
corresponds to a conventional 2D superconducting sample\cite{goldman83} 
(Hg-Xe alloy) and collapses for $T_{KTB}=3.04$~K, $b=9.64$, and 
$z=5.6\pm0.3$ as shown in Fig.~\ref{repacisc}c. 
The apparent agreement between these four diverse samples
suggests that the mean result, $z\simeq 5.7$, may
indeed be universal for superconductors.
However, it was not possible to probe
the universality of the scaling functions, because of  
the non-universal parameter $b$ appearing in $\xi$.

The allure of the KTB transition lies in the simplicity and integrity
of the pair-dissociation paradigm.  It is therefore unsettling that the 
corresponding dynamic paradigm could fail, as signaled above through the
observation of $z\gtrsim 2$.  To better understand this situation, it is
helpful to reconcile the intuitive (``conventional") picture with the
scaling description.  The conventional paradigm intrinsically involves
the presence of dissociated or ``free" vortices, with density $n_f$.  In
a small but finite current, free vortices are independently driven across
the sample, inducing dissipation consistent with the Bardeen-Stephen
theory:  $R\propto n_f$.  In larger currents, non-ohmic effects may
arise from steady state dissociation-recombination processes, although
dissipation is still understood in terms of driven, non-interacting vortices.
To make connection with the scaling approach, we compare the dimensional
statement $n_f\propto \xi^{-2}$, which cannot be affected by dynamical
processes, with the 2D ansatz\cite{ffh} $R\propto \xi^{-z}$.
This leads immediately to the relation $R\propto n_f^{z/2}$, which 
reproduces the conventional relation when $z=2$. Using this 
relationship in the derivation of the conventional signatures, 
one finds $R(T)\propto \exp[-z\sqrt{b/(T/T_{KTB}-1)}]$ and that
$\alpha$ jumps from $z+1$ to 1 at $T_{KTB}$, consistent with the dynamic
scaling results.

The description in terms of a general $z$ now allows for other more
sophisticated dissipation phenomena.  In particular, the observed
value of $z\simeq 5.7$ cannot be mistaken to be 
diffusive, and suggests transport
processes which are highly collaborative in nature.  We speculate that
for superconductors, driven vortices may never be ``free" in the 
conventional sense.  Rather, the dominant transport mechanism may
involve {\it collaborative} dissociation, in which pairs of vortices can unbind
only by exchanging members with neighboring pairs.  In contrast with the
conventional picture, which involves only the two members of the 
dissociating vortex pair, the collaborative dissociation scenario involves
four ``mobile" (as opposed to ``free") vortices:  the dissociating
pair, and the two neighboring vortices with which they recombine.  Assuming
a steady state recombination process analogous to the conventional
one,\cite{goldman83} we now obtain
$\dot{n}_f \sim n_f/\tau \propto n_f^4$, or
$\tau \propto \xi^6$, with a dynamic exponent of 6 rather than 2.
We emphasize that the relatively large observed value of $z$ should
not be attributed to pinning effects, as the phenomenon is observed in
a variety of superconductors and must be regarded as intrinsic.

Finally, we contrast the conclusions of this work with Ref.~\cite{lobb96},
where it was suggested that the upturn of isotherms near $T_{KTB}$, from
ohmic to non-ohmic behavior, may be attributed to electromagnetic
screening, which produces free vortices even below $T_{KTB}$.  Such
effects, which result from the 2D penetration depth becoming smaller
than the sample dimensions, should occur in some samples of finite
thickness, ultimately destroying the KTB transition.  However, in 
analogy with bulk superconductors, the disputed behavior may instead
reflect only the non-ohmic behavior associated with finite currents.
In this scenario, the upturning ohmic curves would correspond to 
$T>T_{KTB}$, and the KTB transition would remain intact.  The 
non-ohmic behavior can be understood as follows.  
On either side of the transition, external 
currents tend to dissociate large vortex pairs through (highly
collaborative) activation processes.  For $T>T_{KTB}$, many large
pairs are already unbound, so that small currents, which
tend to dissociate large pairs, have little effect.  However, beyond
some characteristic current scale ($J_0 \sim T/\xi $), current-induced
dissociation becomes the dominant dissipation mechanism, producing non-ohmic
effects.  It is worth noting that the disputed $I$-$V$ isotherms of
Ref.~\cite{lobb96} are found to scale nicely with the techniques used
here.  (See Fig.~\ref{repacisc}b.)

In summary, a conventional analysis of electronic transport data is 
contrasted with a dynamical scaling analysis for several
types of 2D superconductors. 
The latter approach is designed to
include non-ohmic effects associated with finite currents.  
The conventional analysis obtains
inconclusive results while the dynamical scaling analysis yields a collapse 
of the $I$-$V$ data.
The results show that dynamical critical behavior in 2D superconductors
may be more interesting than previously anticipated.
The scaling generalization of the universal jump of the $I$-$V$ exponent 
$\alpha$ is
from a value of $z+1$ to 1.  Our analysis gives
a dynamical exponent of $z\simeq 5.7$, rather than the
expected value of $z=2$, while a simple dynamical model predicts
$z\simeq 6$.  Our results may pertain more directly to superconductors, where
previous reports of anomalous vortex diffusion have been made,\cite{theron93}
rather than to pure $XY$ models, where there is no indication of 
non-diffusive dynamics.  It remains to be seen how the present 
results are reflected in 2D Josephson junction arrays and granular films.

The authors gratefully acknowledge conversations with S.~E.~Hebboul, 
O.~T.~Valls, B.~I.~Halperin, J.~C.~Garland, S.~M.~Girvin,
M.~Wallin, H.~Gould, P.~Muzikar, and C.~J.~Lobb. 
This work was supported by the Midwest Superconductivity Consortium through 
D.O.E.~Contract No. DE-FG02-90ER45427 and by NSF Grant No. DMR 95-01272. 
Acknowlegement is made to the donors of The Petroleum Research Fund, 
administered by the ACS, for support of this research.


\begin{references}
\bibitem{ktb}J.~M.~Kosterlitz, J.~Phys.~C {\bf 7}, 1046
(1974); J.~M.~Kosterlitz and D.~J.~Thouless, J.~Phys.~C {\bf 6},
1181 (1973); V.~L.~Berezinskii, Sov.~Phys.~JETP {\bf 32}, 493 (1971).
\bibitem{br78}D.~J.~Bishop {\it et al.}, Phys.~Rev.~Lett.~{\bf 40}, 
1727 (1978).
\bibitem{hn79}B.~I.~Halperin and D.~R.~Nelson, J.~Low 
Temp.~Phys.~{\bf 36}, 599 (1979).
\bibitem{huberman79} B.~A.~Huberman {\it et al.}, 
Phys.~Rev.~Lett.~{\bf 43}, 950 (1979).
\bibitem{mertens87}F.~G.~Mertens {\it et al.}, \prl {\bf 59}, 117 (1987).
\bibitem{geer89}R.~Geer {\it et al.}, Phys.~Rev.~Lett.~{\bf 63}, 540 
(1989).
\bibitem{epstein81}K.~Epstein {\it et al.}, Phys. Rev.~Lett.~{\bf 47}, 534 
(1981).
\bibitem{hebard83}A.~F.~Hebard {\it et al.}, Phys.~Rev.~Lett.~{\bf 
50}, 1603 (1983).
\bibitem{goldman83}A.~M.~Kadin {\it et al.}, Phys. Rev.~B {\bf 27}, 
6991 (1983).
\bibitem{resnick81}D.~J.~Resnick {\it et al.}, Phys.~Rev.~Lett.~{\bf 47}, 
1542 (1981).
\bibitem{norton93} D.~P.~Norton {\it et al.}, Phys.~Rev.~B 
{\bf 48}, 6460 (1993).
\bibitem{ktexp}See e.g., D.~H.~Kim,  {\it et al.}, Phys.~Rev.~B {\bf 40}, 
8834 (1989); S.~N.~Artemenko {\it et al.}, JETP Lett. {\bf49}, 654 
(1989); N.-C.~Yeh and 
C.~C.~Tsuei, Phys.~Rev.~B {\bf 39}, 9708 (1989).
\bibitem{minnhagen87}P.~Minnhagen, Rev.~Mod.~Phys.~{\bf 59}, 1001 
(1987).
\bibitem{nelson77}D.~R.~Nelson {\it et al.}, 
Phys.~Rev.~Lett.~{\bf 39}, 1201 (1977).
\bibitem{lobb96} J.~M.~Repaci {\it et al.}, Phys.~Rev.~B {\bf 54}, R9674 
(1996).
\bibitem{rutgers91}S.~Vadlamannati {\it et al.}, Phys.~Rev.~B {\bf 44}, 
7094 (1991).
\bibitem{blatter94}See, e.g., Fig.~29 of G. Blatter {\it et al.}, \rmp {\bf 66}, 1162 (1994).
\bibitem{ffh}D.~S.~Fisher {\it et al.}, Phys. Rev.~B {\bf 43}, 130 (1991).
\bibitem{booth}J.~C.~Booth {\it et al.}, \prl {\bf 77}, 
4438 (1996); K.~Moloni {\it et al.}, \prl {\bf 78}, 3173 (1997), and
to be published; D. Ginsberg {\it et al.}, to be published.  
Related 3D Monte Carlo results include K. H. Lee and D. 
Stroud, \prb {\bf 46}, 5699 (1992); H. Weber and
H. J. Jensen, \prl {\bf 78}, 2620 (1997); J. Lidmar {\it et al.},
(unpublished).  We are not aware of any discussion of non-diffusive
dynamics in 2D superconductors.
\bibitem{wolf79} S.~A.~Wolf {\it et al.}, \prl {\bf 42}, 324 (1979).
\bibitem{note1}In conventional theory,\cite{hn79} an additional 
temperature dependence enters the $T<T_{KTB}$ asymptotic form, which
cannot be accommodated by standard scaling theory.
\bibitem{miu95}L.~Miu {\it et al.}, Phys.~Rev.~B {\bf   52}, 4553 (1995). 
\bibitem{trawick96} M. L. Trawick {\it et al.}, J.~Low Temp.~Phys.~{\bf 
105}, (1996). 
\bibitem{keener97a}C.~D.~Keener {\it et al.}, submitted to
Physica C (1997).
\bibitem{SCdetail} One advantage of this scaling form over the original 
is that it does not stretch out the $y$ scale, only the $x$ scale. 
In addition, the asymptotic behaviors become very simple.
\bibitem{symmdetail}Such symmetry has not been verified theoretically.
One could introduce asymmetry about $T_{KTB}$ through the nonuniversal
parameter $b\rightarrow b_\pm$, as seems most likely.  However, this 
procedure increases the number of fitting parameters, and is not
attempted here.  The assumed symmetry prohibits using
the Minnhagen\protect\cite{minnhagen87} 
form of the resistance. 
\bibitem{theron93}R.~Th\' eron {\it et al.}, \prl {\bf 71}, 1246 (1993).

\end{references}
\end{document}